\documentclass[british]{article}
\usepackage[T1]{fontenc}
\usepackage[latin9]{inputenc}
\usepackage{xcolor}
\usepackage{refstyle}
\usepackage{amsmath}
\usepackage{amsthm}
\usepackage{amssymb}
\usepackage{stmaryrd}
\usepackage{graphicx}

\makeatletter

\AtBeginDocument{\providecommand\figref[1]{\ref{fig:#1}}}

\RS@ifundefined{subsecref}
  {\newref{subsec}{name = \RSsectxt}}
  {}
\RS@ifundefined{thmref}
  {\def\RSthmtxt{theorem~}\newref{thm}{name = \RSthmtxt}}
  {}
\RS@ifundefined{lemref}
  {\def\RSlemtxt{lemma~}\newref{lem}{name = \RSlemtxt}}
  {}

\numberwithin{equation}{section}
\numberwithin{figure}{section}

\makeatother

\usepackage{babel}
\begin{document}

\title{Combining Radon transform and Electrical Capacitance Tomography for
a $2d+1$ imaging device}

\author{Yves Capdeboscq\thanks{Mathematical Institute, Andrew Wiles Building, University of Oxford,
OXFORD OX2 6GG} \and Hrand Mamigonians\footnote{Zedsen Ltd., Chiswick Park, 566 Chiswick High Road, London, W4 5YA}\and Aslam Sulaimalebbe$^{\dagger}$ \and Vahe Tshitoyan$^{\dagger}\footnote{Presently at Lawrence Berkeley National Laboratory, 1 Cyclotron Rd, Berkeley 94720, CA, USA}$}
\maketitle
\begin{abstract}
This paper describes a coplanar non invasive non destructive capacitive
imaging device. We first introduce a mathematical model for its output,
and discuss some of its theoretical capabilities. We show that the
data obtained from this device can be interpreted as a weighted Radon
transform of the electrical permittivity of the measured object near
its surface. Image reconstructions from experimental data provide 
good surface resolution as well as short depth imaging, making the 
apparatus a $2d+1$ imager. The quality
of the images leads us to expect that excellent results can be
delivered by \emph{ad-hoc} optimized inversion formulas. There are
also interesting, yet unexplored, theoretical questions on imaging
that this sensor will allow to test. 
\end{abstract}

\section{Introduction}

The imaging capability of electrical capacity sensors -- sometimes
called dielectrometry -- has been discussed in connection with applications
to a variety of domains \cite{1288505,7954985,7956144}, from chemical
engineering \cite{RIMPILAINEN2012220} to medical imaging \cite{7932938}.
It is very closely related to Electrical Impedance Tomography (EIT)
in terms of the partial differential equation used in the associated
quasi-static model \cite{webster1990electrical,0266-5611-18-6-201,holder2005electrical,0266-5611-25-12-123011,Adler2015}.
Several features are nevertheless specific to this modality: in particular,
physical contact is not required between the sample and the measurement
device. Furthermore, measurements can be performed on one side only
\cite{0960-1317-25-7-075025,1703487,1288505}. The design of electrodes
for Electrical Capacitance Tomography (ECT) and EIT is different.
Inter-digital, periodic, grating sensors or the multiplexing device
described in this note are line sensors \cite{1288505,1703487,7954985}:
their output cannot be approximated as pointwise data. Volumetric
capacitive measurements, using planar sensors of smaller diameter,
have been discussed and analysed in many articles \cite{7956144,02602281011010772,Ye201313542575675}.
The purpose of these studies is to detect the presence of objects
at a significant depth. Here the imaging device under consideration
is focused on retrieving near-field details, rather than deep coarse
features. 

\begin{figure*}
\centering{}\resizebox{0.8\columnwidth}{!}{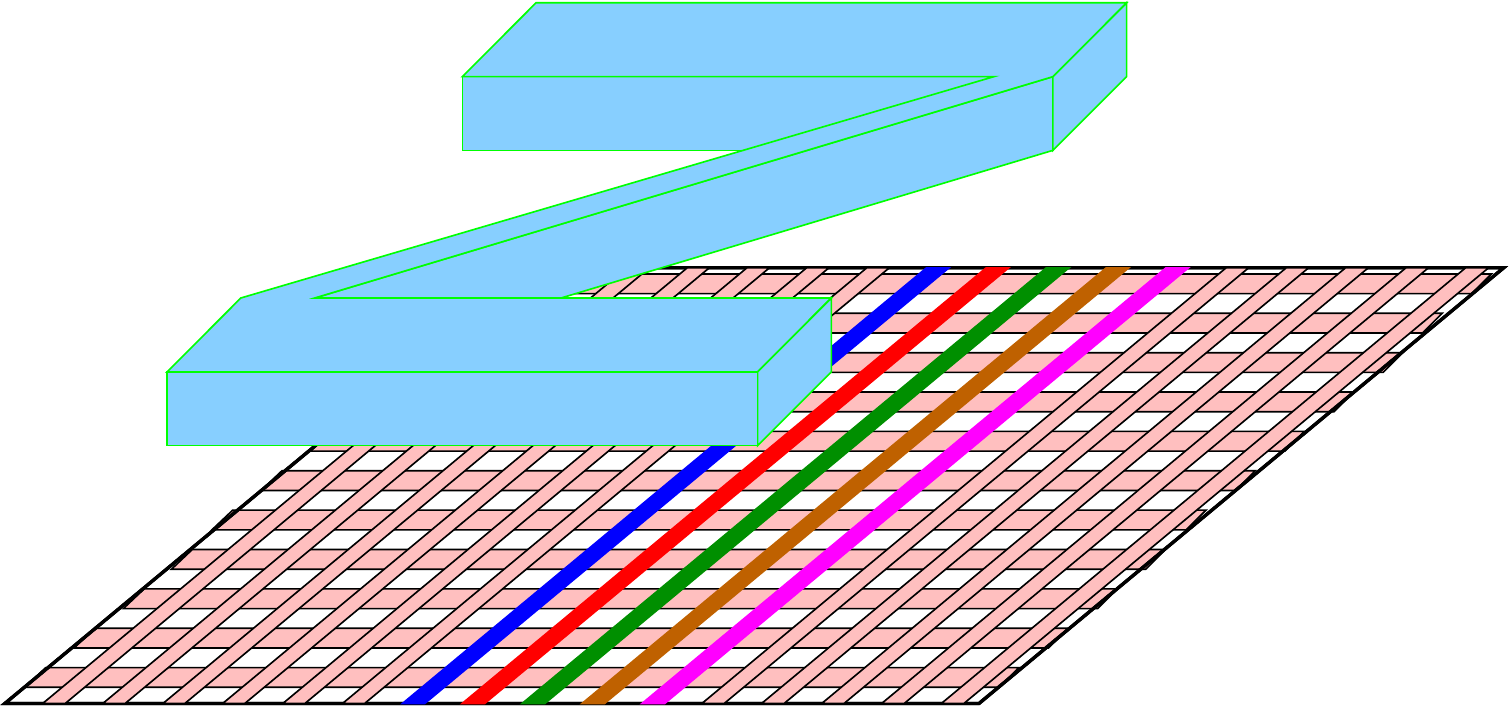}\caption{\label{fig:first} Sketch of the imaging setup.}
\end{figure*}

\begin{figure}
\begin{centering}
\includegraphics[width=0.95\columnwidth]{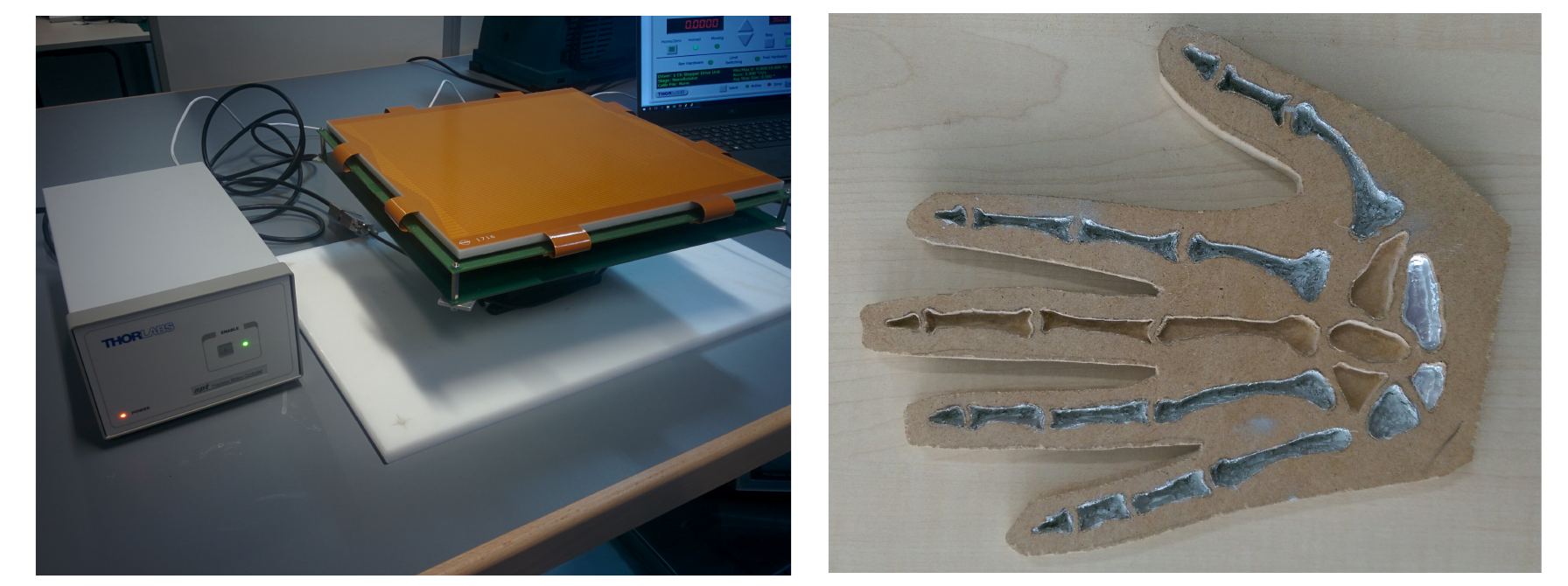}
\par\end{centering}
\centering{}\caption{\label{fig:Kenny} The hand model (Kenny's) is approximately $200$mm long. The sensor array has $55\times56$ sensor lines $d=2.5$mm apart, covering
a $280\times280$mm square. There were $36,90$ or 180 rotational increments ($1,2$ or $5$ degrees each time). The hand was suspended $2$mm above the sensor array, avoiding contact. Silver ink was used to paint most of the trenches, whose lowest point is $5$mm above the sensor. The trenches in the middle finger and some trenches on the palm were intentionally left empty.}
\end{figure}

The imaging device under consideration is flat: it is represented
schematically in \figref{first}, and physically in \figref{Kenny}.
It has been developed by Zedsen Limited. The dielectric material (a
light blue Z, for illustration purposes) is held above the apparatus.
The sensor is composed of a series of narrow electrodes \cite{patent:8994383},
which can act as either receiver or transmitter. In \figref{first},
assuming that the leftmost (blue) electrode is the emitter, either
the red, green, brown or magenta electrode is used as a receiver,
while all others are grounded. The transmitting electrode sends a
pulse for a short duration. The resulting voltage is measured on the
receiving electrodes at several time steps. Thus, for a given grid
of $2n+1$ electrodes, we obtain ${\color{red}2n}+{\color{green}2n-1}+{\color{brown}2n-2}+{\color{purple}2n-3}$
measures, at a few time points. The imaging device is equipped with
a second set of $2m+1$ electrodes placed orthogonally to the first
set, leading to another $8m-6$ measurements. Cross measurements are
also possible, e.g. transmitting from an horizontal electrode and
measuring on a vertical one or vice-versa, leading to $\left(2n+1\right)\times\left(2m+1\right)$
measurements \footnote{Strictly speaking, $2\times\left(2n+1\right)\times\left(2m+1\right)$
measurements, but only $\left(2n+1\right)\times\left(2m+1\right)$
are independent by the reciprocity Theorem, as explained in the sequel.}. 

\begin{figure}
\centering{}\resizebox{0.8\columnwidth}{!}{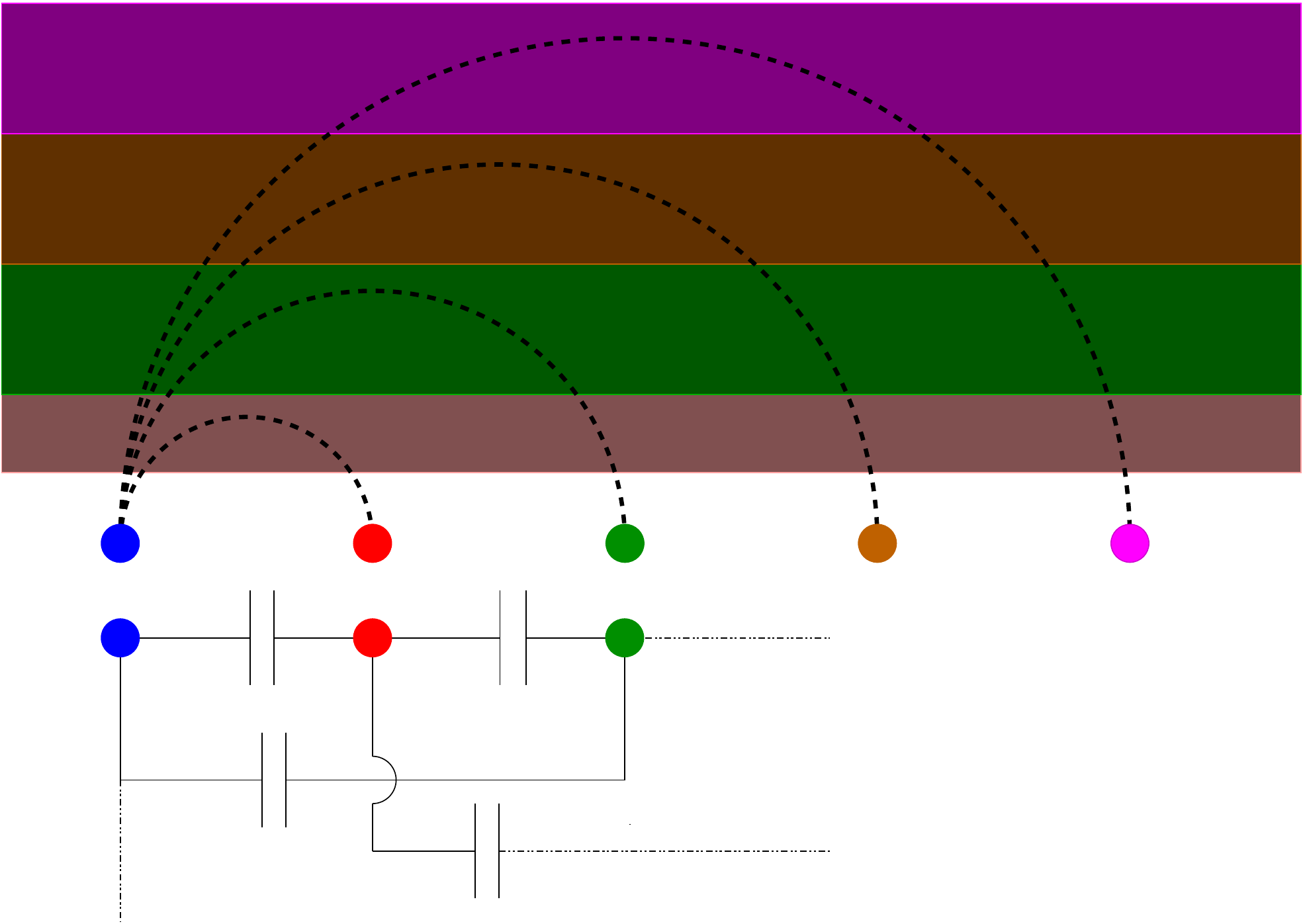}\caption{\label{fig:cross} Schematic cross section, fringing field and capacitive
network interpretations.}
\end{figure}

The rationale for such measurements is based on two dimensional problem
reduction considerations. Consider a vertical cross-section of the
problem as described in \figref{cross}. If a long vertical homogeneous
dielectric is placed on the sensor, the problem becomes approximately
two dimensional, and the red, green, brown and magenta electrodes
measure the output at an increasing distance. Modelling the system
as an electric circuit, which can be justified for very contrasted
materials (in the case of electrical impedance tomography, see \cite{0266-5611-25-12-123011,0266-5611-24-3-035013,0266-5611-18-6-201})
this leads to the determination of up to $10$ capacitances. Alternatively,
in the case of a weakly contrasted material, viewing the system as
a perturbation of the free space, the fringing field zone of dependence
-- the area where the gradient of the electrostatic potential is
large enough to have an effect -- is often modelled as semi-discs.
As the distance between the source and the receiver increases, the
data collected by the sensor depends on the permittivity of deeper
and deeper layers. Assuming that the medium is suitably homogeneous,
other planar ECT sensors have been shown to provide some depth capacitive
data, interdigital devices in particular \cite{IGREJA2011392,DIAS201795};
assuming that the electric field decays so fast that its amplitude
is negligible after the fourth electrodes, similar heuristic approximations
could be devised in this case as well. 

\begin{figure}
\centering{}\resizebox{0.8\columnwidth}{!}{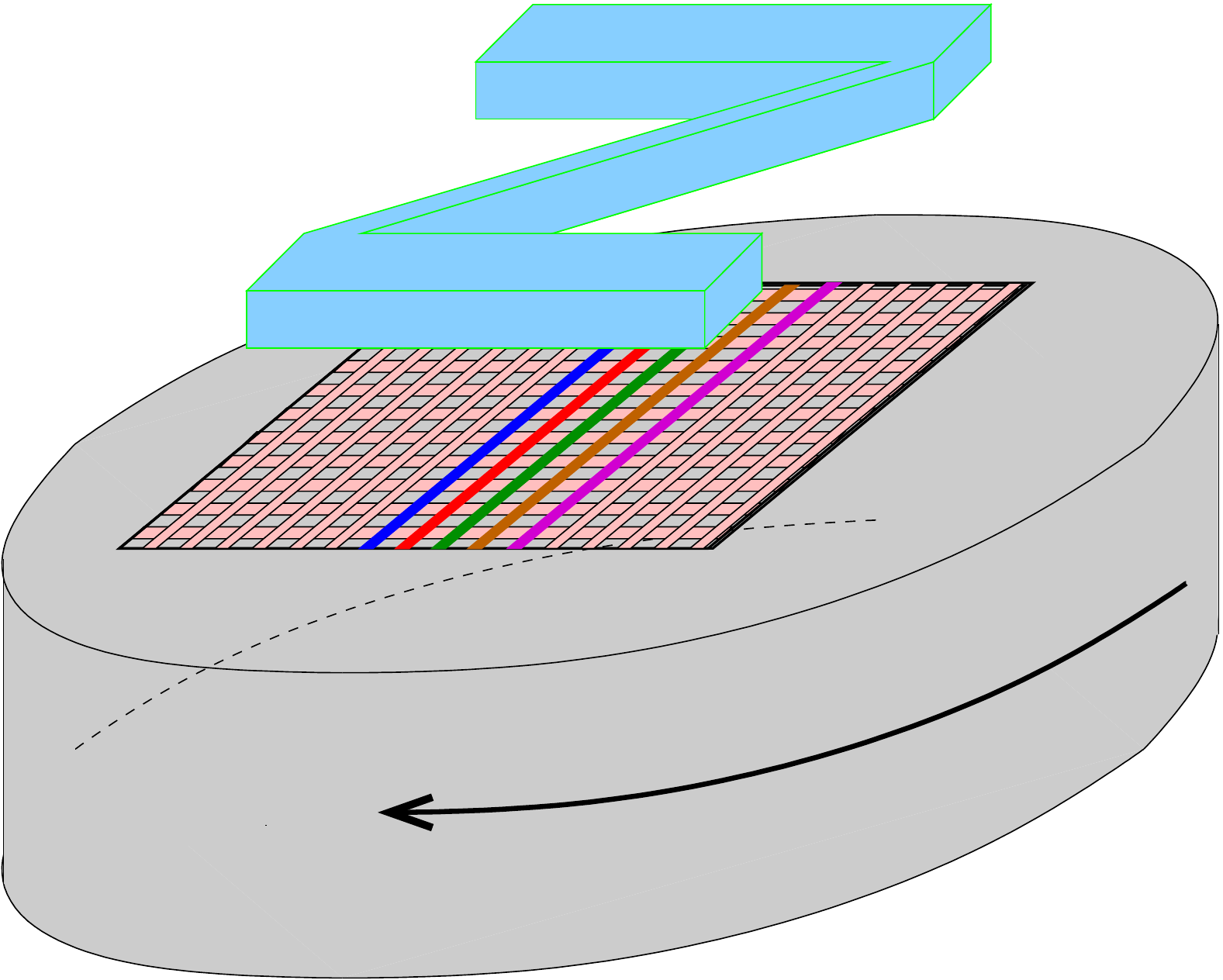}\caption{\label{fig:rotate} The sensor with rotational support. }
\end{figure}

There are limitations to this approach. An object placed diagonally
provides the same output as a horizontal one with an adequate cross
section, since the output are summed along lines: these data sets
are linear in $n$ and $m$. The cross sectional measurements, from
vertical to horizontal electrodes or vice-versa, do provide more localised
information, with potentially $\left(2n+1\right)\times\left(2m+1\right)$
independent data points. However, depth dependent information is then
only available, indirectly, by means of different time points. The
theoretical framework to take advantage of such data is under-developed
at this time, and preliminary computations in model cases tend to
indicate that the dependence on time difference data could be logarithmic. 
Collating the localized
horizontal data and vertical or horizontal depth measurements does
not scale correctly (with respect to the interdigital distance) to
provide multi-layered data, but for coarse horizontal grids. One feature
of this sensor device is that its horizontal resolution, namely, the
distance between the line electrodes, can be adapted to the application
at hand. This makes it different from interdigital sensors, where
the distance between the gratings determines a wavelength, which is
of course related to a spatial resolution, but indirectly. To address
these limitations, the dielectrometric imaging device has been mounted
on a rotating support, as represented in \figref{rotate} and pictured
in \figref{Kenny}. 

The dielectric to be imaged is static, while the sensor array is placed
on a turntable, rotating by with fixed angular increments of $\pi/p$
radians. As the data are non-directional, all non redundant data are
collected after $p$ angular steps. All together, the data collected
is an array of size $10p\times\left(\frac{2}{5}n\times m+n+m-1\right)$, 
for every time-point.

\section{Mathematical Modelling}

\subsection{\label{subsec:mathcap}What the non-rotational output data represents }

The preferred model used by practitioners is (quasi-)electrostatic
\cite{02602281011010772,1288505,1703487,7932938,Banasiak2010241,Soleimani20051987,Tholin-Chittenden20178059,Ye201313542575675}.
We write $\underline{x}=\left(x_{1},x_{2},z\right)$ the Cartesian
coordinates of any point. The domain under consideration $\Omega$
is outside of the measurement device, where the permittivity is given
by $\varepsilon_{0}\left(\underline{x}\right)$ when no object to
be measured is present. Above the sensor array, $z\geq0$, the medium
is considered homogeneous, $\varepsilon_{0}\left(\underline{x}\right)=\varepsilon_{0}$.
At $z=0$, an insulating board is placed, shielding the electronic
equipment and power sources (located below) from interferences with
the measurements. Our domain is therefore 
\[
\Omega=\mathbb{R}^{2}\times\left(0,\infty\right)\setminus\cup_{i=1}^{N}I_{i},
\]
where $I_{1},\ldots,I_{N}$ are the electrodes, and $N=2\left(n+m+1\right)$
is the total number of electrodes, horizontal and vertical. The quasi-static
approximation corresponds to the absence of current in the measured
object $O$, thus behaving like a perfect dielectric \cite[II.2.a]{zbMATH01388032}.
This means that non-zero charges are only on the surface of the (perfectly
conducting) electrodes, $I_{1,},\ldots,I_{N}$. The total charge on
the electrodes $I_{i}$ is 
\[
q_{i}=\int_{\partial I_{i}}\rho_{i}\left(\underline{x}^{\prime}\right)\text{d}\sigma\left(\underline{x}^{\prime}\right),
\]
 where $\rho_{i}\left(\underline{x}^{\prime}\right)$ is the surface
charge density of $I_{i}$ (and $\sigma$ is the surface measure of
$\partial I$). If we write $U_{i}$ as the (constant) potential on
$I_{i}$, we see that the electrostatic potential $u^{0}$ (in absence
of any object $O$ to measure) satisfies 
\begin{equation}
\begin{cases}
\text{div}\left(\varepsilon_{0}\nabla u^{0}\right)=0 & \text{ in }\text{\ensuremath{\Omega},}\\
\left.\begin{array}{lr}
u^{0}\left(\underline{x}^{\prime}\right) & =U_{i}\\
\left(\varepsilon_{0}\nabla u^{0}\cdot\nu\right)\left(\underline{x}^{\prime}\right) & =\rho_{i}\left(\underline{x}^{\prime}\right)
\end{array}\right\}  & \text{on }\partial I_{i}\text{ for }i=1,\ldots,N,\\
\varepsilon_{0}\partial_{z}u^{0}(x,y,0)=0,\\
{\displaystyle \lim_{\left|\underline{x}\right|\to\infty}}u^{0}\left(\underline{x}\right)=0,
\end{cases}\label{eq:CapaZero}
\end{equation}
where on each surface $\partial I_{i},$ $\nu$ represent the outgoing
normal. Naturally, one cannot impose both all charges and all voltages,
as it would be an overdetermined problem. The (mathematical) capacity
$C^{0}:\mathbb{R}^{N}\to\mathbb{R}^{N}$ is the map linking $\left(U_{1},\ldots,U_{N}\right)\to\left(q_{1},\ldots,q_{N}\right)$
via (\ref{eq:CapaZero}); it is well defined \cite[II.2.a]{zbMATH01388032}.
It is linear, by the superposition principle, and therefore can be
represented as a $N\times N$ matrix 
\[
\underline{\underline{C^{0}}}=\left(C_{ij}^{0}\right)_{1\leq i,j\leq N},\text{ so that }q_{i}=\sum_{j=1}^{N}C_{ij}^{0}U_{j},
\]
where 
\begin{equation}
C_{ij}^{0}=\int_{\partial I_{i}}\varepsilon_{0}\nabla u_{j}^{0}\cdot\nu\left(\underline{x}^{\prime}\right)\text{d}\sigma\left(\underline{x}^{\prime}\right)\label{eq:C0ij}
\end{equation}
and $u_{j}^{0}$ is the solution tending to zero at infinity such
that 
\begin{equation}
\text{div}\left(\varepsilon_{0}\nabla u_{j}^{0}\right)=0\text{ in }\Omega,\label{eq:DefUoJ}
\end{equation}
and such that $u_{0}^{j}=1$ on $I_{j}$, and $u_{0}^{j}=0$ on $I_{k}$
if $k\neq j$. In presence of an object $O\subset\Omega$ of permittivity
is $\varepsilon_{r}\left(\underline{x}\right)\varepsilon_{0}$, the
total permittivity becomes 
\[
\varepsilon=\epsilon_{0}\begin{cases}
1 & \text{ outside }O,\\
\varepsilon_{r}(x) & \text{ in }O.
\end{cases}
\]
Extending the definition of $\epsilon_{r}$ by $1$ outside $O$ the
permittivity becomes $\varepsilon_{0}\varepsilon_{r}$ provided the
object does not touch the electrodes, the capacity matrix is given
by 
\begin{equation}
C_{ij}=\int_{\partial I_{i}}\varepsilon_{0}\nabla u_{j}\cdot\nu\left(\underline{x}^{\prime}\right)\text{d}\sigma\left(\underline{x}^{\prime}\right),\label{eq:defCij}
\end{equation}
where $u_{j}$ is the unique weak solution tending to zero at infinity
such that $u_{j}=u_{j}^{0}$ on the electrodes and 
\[
\text{div}\left(\varepsilon_{0}\varepsilon_{r}\left(\underline{x}\right)\nabla u_{j}\right)=0\text{ in }\Omega,
\]
which means that for any smooth $\phi$ compactly supported in $\Omega$,
\[
\int_{\mathbb{R}^{2}\times\left[0,\infty\right)\setminus\cup_{i=1}^{N}I_{i}}\varepsilon_{0}\varepsilon\left(\underline{x}\right)\nabla u_{j}\cdot\nabla\phi\,\text{d}x\text{d}y\text{d}z=0.
\]
An integration by parts shows that 
\begin{align*}
C_{ij} & =\int_{\Omega}\varepsilon_{0}\varepsilon_{r}\nabla u_{j}\cdot\nabla u_{i}\text{d}x\text{d}y\text{d}z,\\
\text{ and }C_{ij} & =\int_{\Omega}\varepsilon_{0}\varepsilon_{r}\nabla u_{j}\cdot\nabla u_{i}^{0}\text{d}x\text{d}y\text{d}z.
\end{align*}
The first formula shows that the Capacity matrix is symmetric --
this is the Reciprocity Theorem. The second formula shows that relative
capacity $\Delta C$ is 
\begin{equation}
\Delta C_{ij}=C_{ij}-C_{ij}^{0}=\int_{O}\varepsilon_{0}\left(\varepsilon_{r}-1\right)\nabla u_{j}\cdot\nabla u_{i}^{0}\text{d}x\text{d}y\text{d}z,\label{eq:substracted-data}
\end{equation}
an integral over the object to be imaged. 
Applying the Hopf's Maximum principle, we deduce from \eqref{defCij} that for any $i\neq j$, $C_{ij}<0$.
Thus no information is lost by recording only the moduli of $C_{ij}$. 

We denote by $H_{i,k}$ (resp. $V_{i,k}$) the output data obtained
obtained between the horizontal (resp. vertical) electrode $i$ and
$k$-th electrode to its right. We denote by $X_{ij}$ the output
obtained when a horizontal electrode $i$ is used as an emitter and
a vertical electrode $j$ is used as receiver (or vice-versa, which
is equivalent). Keeping the colour coding of \figref{first},
the data given by the imaging device are entries of the relative capacity
matrix $\Delta C$, namely, for the first for $2n+1$ lines and columns,
\[
\Delta C=\left[\begin{array}{ccccc}
\Delta C_{11} & {\color{red}H_{1,1}} & {\color{green}H_{12}} & H_{{\color{black}1,3}}\\
{\color{red}H_{1,1}} & \ddots & \ddots & \ddots & {\color{black}H_{2n-2,3}}\\
{\color{green}\vdots} & \ddots & \ddots & \ddots & {\color{green}H_{2n-1,2}}\\
{\color{purple}H_{{\color{black}1,4}}} &  & \ddots & \ddots & {\color{red}H_{2n,1}}\\
 & {\color{black}{\color{purple}H_{2n-3,4}}} & \ldots & {\color{red}H_{2n,1}} & \Delta C_{2n+1,2n+1}
\end{array}\right],
\]
for the last $2m+1$ lines and columns,
\[
\Delta C=\left[\begin{array}{ccccc}
\Delta C_{2n+2,2n+2} & {\color{red}V_{1,1}} & {\color{green}V_{1,2}} & V_{{\color{black}1,3}}\\
{\color{red}V_{1,1}} & \ddots & \ddots & \ddots & {\color{black}V_{2m-2,3}}\\
\vdots & \ddots & \ddots & \ddots & {\color{green}V_{2m-1,2}}\\
{\color{purple}V_{{\color{black}1,4}}} & \ddots & \ddots & \ddots & {\color{red}V_{2m,1}}\\
 & {\color{black}{\color{purple}V_{2m-3,4}}} & \ldots & {\color{red}V_{2m,1}} & \Delta C_{N,N}
\end{array}\right],
\]
whereas for the off-diagonal blocks, 
\[
\Delta C_{i,j}=X_{i,j}.
\]
Note that the self-capacitance, $C_{ii}=\int_{\Omega}\varepsilon_{0}\varepsilon_{r}\nabla u_{i}\cdot\nabla u_{i}$
isn't measured. In two dimension, one can show that $C_{ii}=-\sum_{j\neq i}C_{ij}$,
making this data redundant. This is also the case for interior problems
in both two and three dimensions. However, in three dimensions, for
exterior problem -- such as the case at hand -- it is always the
case that \cite[II.5.2b]{zbMATH01388032}
\[
C_{ii}>-\sum_{j\neq i}C_{ij}.
\]
This fact makes any three dimensional planar sensor different from its
two dimensional cross sectional approximations, and measurements performed
within a cylinder geometry. The other potentially inaccessible data,
are longer range interaction between horizontal and vertical sensors,
i.e., more than four gaps aside. Such data can be collected, however
the amplitude of such signal decays rapidly. 

\subsection{\label{subsec:Born}The Linear--Born--Small Amplitude Approximation}

The available data is therefore 
\begin{equation}
\Delta C_{ij}=\int_{\mathbb{R}^{3}}\varepsilon_{0}\left(\varepsilon_{r}-1\right)\nabla u_{j}\cdot\nabla u_{i}^{0}\text{d}x\text{d}y\text{d}z,\label{eq:DeltaData}
\end{equation}
which depends non-linearly on $\varepsilon_{r}$ by means of the term
$\nabla u_{j}$. A frequently adopted approximation is to linearise
this problem, namely, to replace $\nabla u_{j}$ by $\nabla u_{j}^{0}$
in this formula \cite{ISI:A1984TF31700004,MR1036240,Adler2015}. In
Electromagnetism, this is the so-called Born approximation. We can
write the Partial Differential Equation satisfied by $u_{j}$ under
the form 
\[
\text{div}\left(\varepsilon_{0}\nabla u_{j}\right)=-\text{div}\left(\varepsilon_{0}\left(\epsilon_{r}-1\right)\nabla u_{j}\right)\text{ in }\Omega,
\]
and assuming that $\left(\epsilon_{r}-1\right)$ is small enough,
this is approximately 
\[
\text{div}\left(\varepsilon_{0}\nabla u_{j}\right)\approx-\text{div}\left(\varepsilon_{0}\left(\epsilon_{r}-1\right)\nabla u_{j}^{0}\right)\text{ in }\Omega.
\]
The same integration by parts as before then gives 
\[
\Delta C_{ij}\approx\int_{\mathbb{R}^{3}}\varepsilon_{0}\left(\varepsilon_{r}-1\right)\nabla u_{j}^{0}\cdot\nabla u_{i}^{0}\text{d}x\text{d}y\text{d}z,
\]
which has the added benefit of depending linearly on $\varepsilon_{r}$. 

Assuming that the object is placed in the middle of the sensor, that
is, there is a number of uncovered electrodes on all sides of the
support of $O$ of $\left(\varepsilon_{r}-1\right)$, then from within
$O$, the array ``appears'' infinitely extended in all direction.
Thus, for $1\leq k\leq4$ we have,
\begin{align*}
H_{i,k} & \approx\int_{\mathbb{R}^{3}}\varepsilon_{0}\left(\varepsilon_{r}\left(\underline{x}\right)-1\right)\left(\nabla u_{n+1}^{0}\cdot\nabla u_{n+1+k}^{0}\right)\left(\underline{x}-a_{i}\left(1,0,0\right)\right)\text{d}x\text{d}y\text{d}z,\\
 & \approx\int_{\mathbb{R}^{3}}\varepsilon_{0}\left(\varepsilon_{r}\left(\underline{x}\right)-1\right)\psi_{k}\left(\underline{x}-x_{i}\left(1,0,0\right)\right)\text{d}x\text{d}y\text{d}z,
\end{align*}
with $a_{i}=\left(i-n-1\right)d\in\left\llbracket -nd,\left(n-k\right)d\right\rrbracket $
and 
\[
\psi_{k}=\nabla u_{n+1}^{0}\cdot\nabla u_{n+1+k}^{0}.
\]
Similarly, there holds 
\[
V_{j,k}\approx\int_{\mathbb{R}^{3}}\varepsilon_{0}\left(\varepsilon_{r}\left(\underline{x}\right)-1\right)\phi_{k}\left(\underline{x}-b_{j}\left(0,1,0\right)\right)\text{d}x\text{d}y\text{d}z,
\]
with $b_{j}=\left(j+m-N\right)d\in\left\llbracket -md,\left(m-k\right)d\right\rrbracket $
and 
\[
\phi_{k}=\nabla u_{N-m}^{0}\cdot\nabla u_{N-m+k}^{0}.
\]
The cross data can be written in the same way
\[
X_{i,j}\approx\int_{\mathbb{R}^{3}}\varepsilon_{0}\left(\varepsilon_{r}\left(\underline{x}\right)-1\right)\xi\left(\underline{x}-\left(a_{i},b_{j},0\right)\right)\text{d}x\text{d}y\text{d}z,
\]
with $(a_{i},b_{j})\in\left\llbracket -nd,nd\right\rrbracket \times\left\llbracket -md,md\right\rrbracket $,
and 
\[
\xi=\nabla u_{n+1}^{0}\cdot\nabla u_{N-m+k}^{0}.
\]

The gradient fields decay (exponentially fast) away from the electrodes,
and machine precision is $0.7\%$ (the input voltage is at 140V and
the sensor returns integer values of voltages) thus it is natural
to assume that the $\psi_{k},$ and $\phi_{k}$, $1\leq k\leq4$ and
$\xi$ are supported near the origin. In other words, the middle of
central horizontal and vertical electrodes are above the origin. 

A further simplification consists in not taking into account the presence
of the second layer of (vertical) electrodes, physically located below
the horizontal layer, as they are not located between the electrodes
and the object to the imaged. This means that, at leading order $\psi_{k}$
depends on $x$ and $z$ only. Thus 
\begin{equation}
H_{i,k}\approx\int_{\mathbb{R}^{3}}\varepsilon_{0}\left(\varepsilon_{r}\left(\underline{x}\right)-1\right)\psi_{k}\left(x-x_{i},z\right)\text{d}x\text{d}z\text{d}y,\label{eq:Hnonrot}
\end{equation}

\begin{figure}
\begin{centering}
\includegraphics[width=0.33\columnwidth]{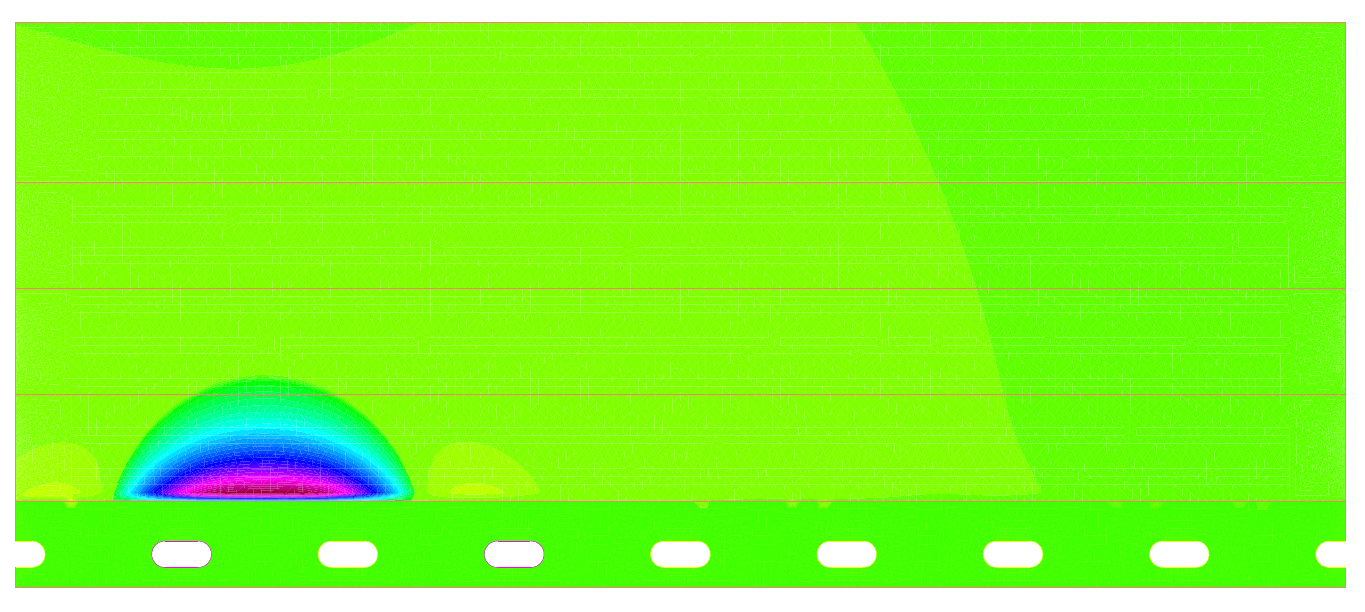}\includegraphics[width=0.33\columnwidth]{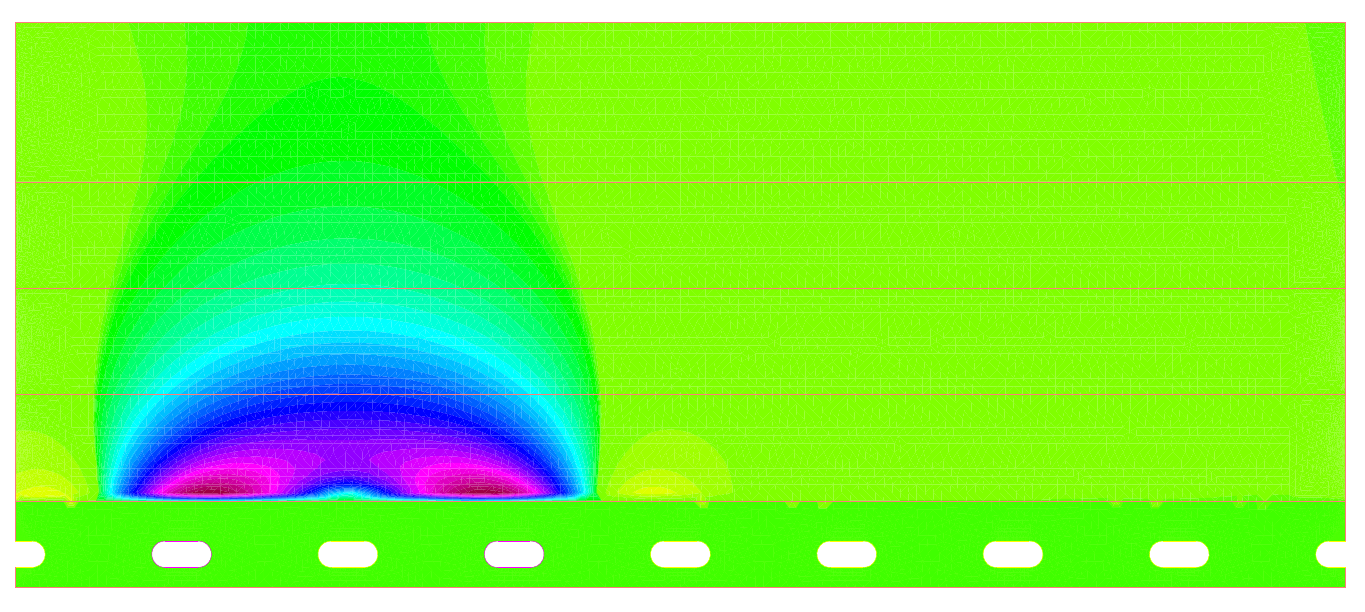}\includegraphics[width=0.33\columnwidth]{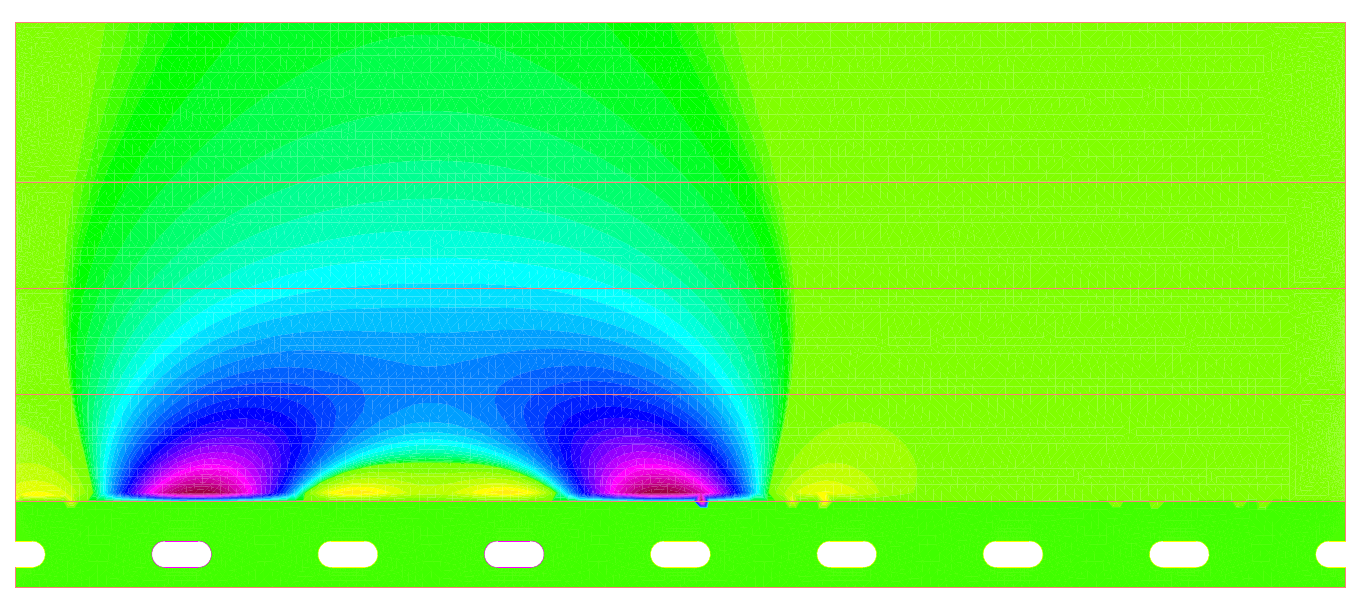}
\par\end{centering}
\centering{}\caption{\label{fig:Psik} The first three horizontal filters $\psi_{1},\psi_{2}$
and $\psi_{3}$. The four horizontal lines correspond to $d$, $3d$,
$5d$ and $7d$, where $d$ is the distance between the electrodes.
Since the object to be imaged is located at distance $d$ (the first
line), $\psi_{1},\psi_{2}$ and $\psi_{3}$ have been set to zero
below that line. They are normalized so that $\max\psi_{k}=1$ for
each $k$. Warmer colours represent higher values, followed by blue,
then green around $0$ and yellow for negative values. The computation
was made on a larger domain, with $14$ electrodes, with Freefem++\cite{FreeFem},
on a fine mesh, using $P2$ elements.}
\end{figure}

A FEM simulated plot of $\psi_{1},\psi_{2}$ and $\psi_{3}$, performed
using Freefem++\cite{FreeFem} is given in \figref{Psik}. We represented
on the graphics the vertical distance $d$, $3d$, $5d$ and $7d$,
where $d$ is the inter-electrode distance. The bottom of the object
we measure is located at distance $d$, therefore we truncated the
support of the weights $\psi_{k}$, $k=1,2,3$, namely we represented
instead $\left(x_{1},x_{2}\right)\to c_{k}\psi_{k}(x_{1,}x_{2})\mathbf{1}_{x_{2}\geq d}$,
where $c_{k}$ is a scaling constant so that $\max\psi_{k}=1$ for
each $k=1,2,3$. Warmer (purple) colours represent the highest values,
followed by shades of blue, then green near $0$ and yellow for negative
values. The logic of the heuristic argument given in the introduction
is roughly satisfied. The area where $\psi_{1}$ takes purple/blue
values is closer $x_{2}=0$ than the corresponding area for $\psi_{2}$,
which is itself lower than the corresponding area for $\psi_{3}$.
However, the weight $\psi_{k}$ are far from uniform, and
their amplitudes decay with distance. 

\begin{figure}
\begin{centering}
\includegraphics[width=0.33\columnwidth]{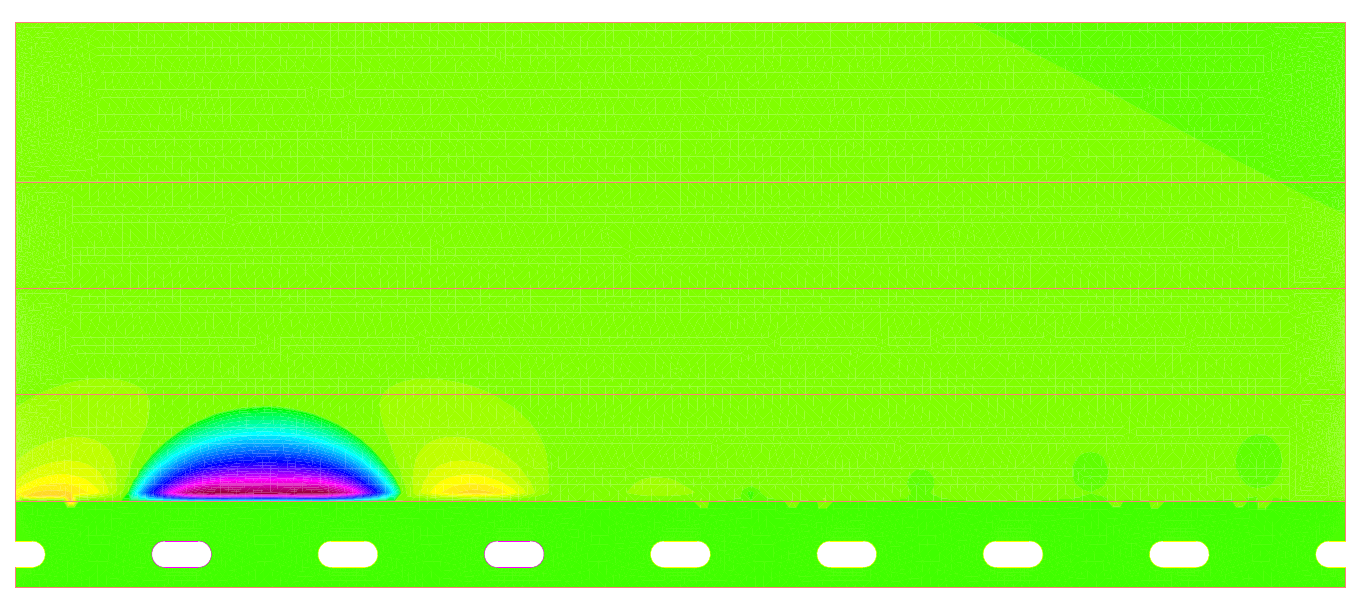}\includegraphics[width=0.33\columnwidth]{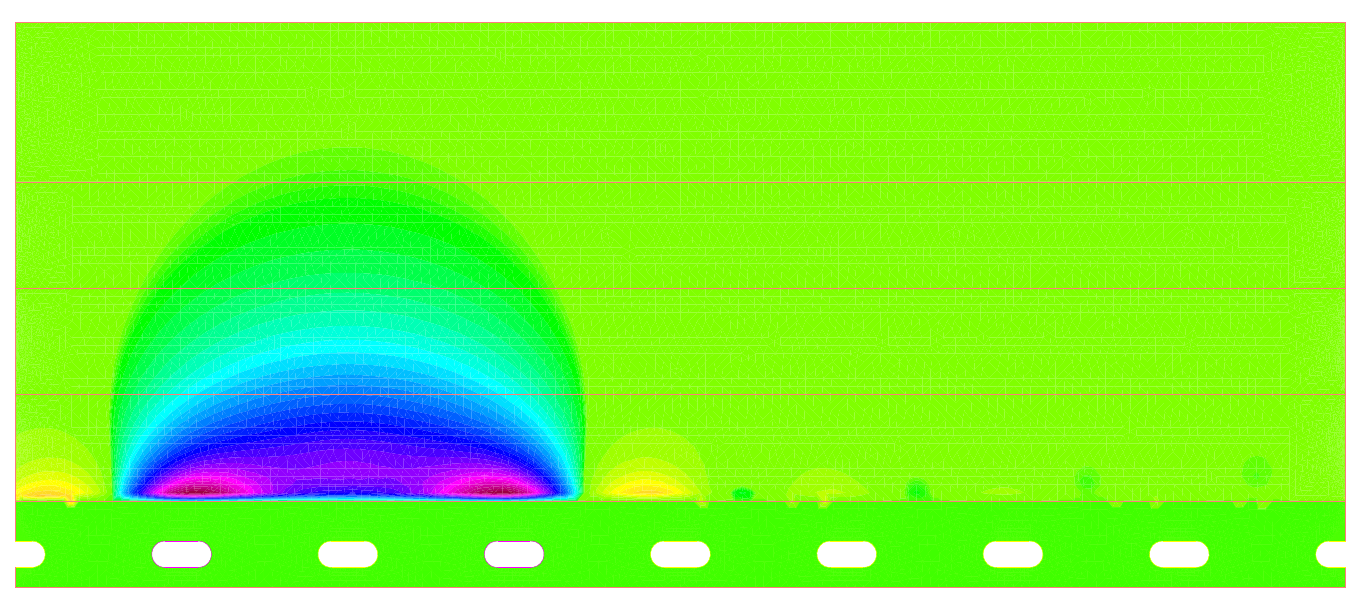}\includegraphics[width=0.33\columnwidth]{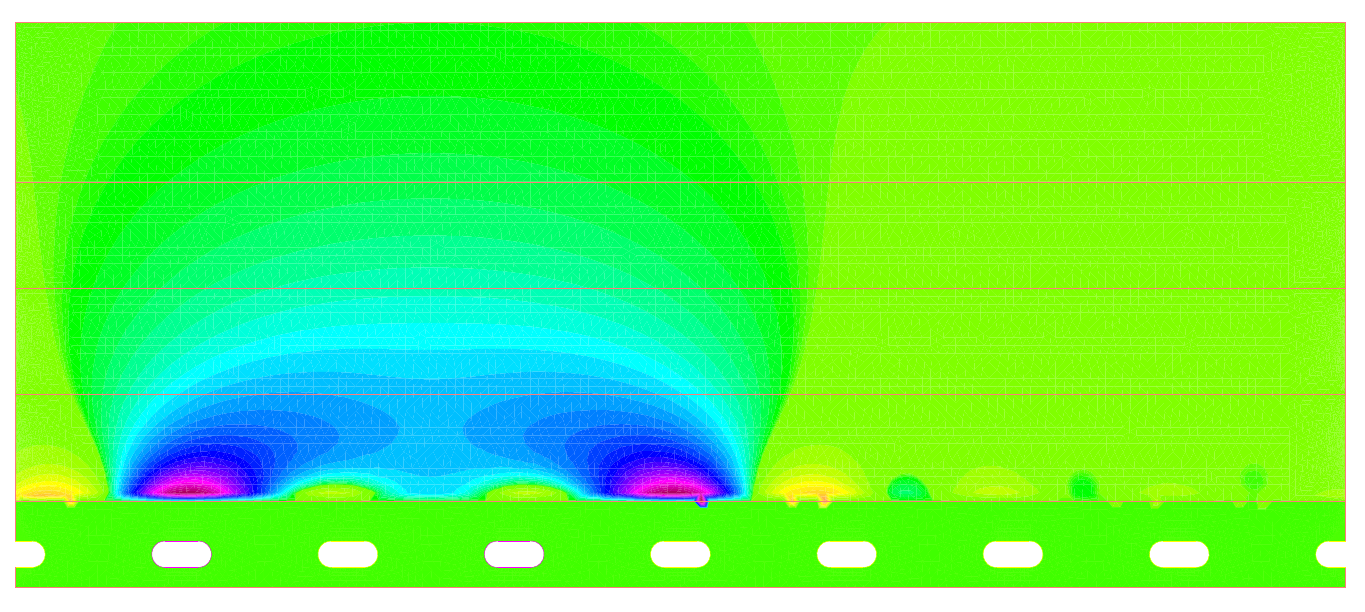}
\par\end{centering}
\centering{}\caption{\label{fig:APsik}The first three horizontal approximate horizontal
filters $\tilde{\psi}_{1},\tilde{\psi}_{2}$ and $\tilde{\psi}_{3}$.
Computed with $N=16,$and $\epsilon_{0}=d/10.$}
\end{figure}

It is worth noting that, for algorithm testing purposes, it is possible
to construct closed form qualitatively satisfactory approximation
formulas for $\psi_{k}$. The advantage of such approximations is
that they do not require the use of a finite element solver. We remind
the reader that the Green function for a one dimensional periodic
array of period $1$ in the direction $x_{1}$ is given by \cite{zbMATH05162228,zbMATH00911725}
\begin{equation}
G(x_{1},x_{2})=-\frac{1}{4\pi}\log\left(\sinh\left(\pi x_{2}\right)^{2}+\sinh\left(\pi x_{1}\right)^{2}\right)+\frac{1}{2}\left|x_{2}\right|.\label{eq:formulaG}
\end{equation}
 It satisfies, in the sense of distributions 
\[
-\Delta G=\sum_{n\in\mathbb{Z}}\delta\left((x_{1},x_{2})-(n,0)\right).
\]
 Given a small parameter $\epsilon_{0}$, we can construct a function
$f_{N}$ such that 
\[
\Delta f_{N}=0\text{ for }x_{2}\ge\epsilon_{0},\,f\left(0,\epsilon_{0}\right)=1\text{ and }f\left(n,\epsilon_{0}\right)=0\text{ for all }n=\left\{ -N,\ldots,-1,1,\ldots,N\right\} ,
\]
by finding the $N+1$ coefficients $\alpha_{0},\ldots,\alpha_{N}$
solving the linear system 
\begin{equation}
M_{N}\left[\begin{array}{c}
\alpha_{0}\\
\vdots\\
\vdots\\
\alpha_{N}
\end{array}\right]=\left[\begin{array}{c}
1\\
0\\
\vdots\\
0
\end{array}\right],\label{eq:linearanal}
\end{equation}
where for all $1\leq i,j\leq N+1$,
\[
M_{ij}=G(\frac{i-1}{j},\epsilon_{0}).
\]
 The function $f_{N}$ is then given by 
\[
f_{N}\left(x,y\right)=\sum_{k=0}^{N}\alpha_{k}G\left(\frac{x_{1}}{k+1},\frac{x_{2}}{k+1}\right).
\]
An approximation for $\psi_{k}$ is then 
\[
\tilde{\psi}_{k}(x,y)=-\nabla f_{N}\left(x,y\right)\cdot\nabla f_{N}\left(x-k,y\right).
\]
This approximates $u^{0}$, the background solution of the capacity
problem (\ref{eq:CapaZero}), by a periodic solution in $x_{1}$ of
large period $N$, which cancels periodically with period $1$ along
the line $x_{2}=\epsilon_{0}$ when $x_{1}$ is an integer and not
a multiple of $N$. Provided that $N$ is chosen large enough, $f_{N}$
is locally qualitatively similar to $u$. One example is represented
in \figref{APsik}. Naturally, because $G$ given by \ref{eq:formulaG}
does not contain any information about the shape of the electrodes,
they do not correspond precisely to \figref{Psik}, but they
capture the main features of these weights. Regarding the solvability
of system (\ref{eq:linearanal}), notice that after one step of Gaussian
elimination, the $N$-by-$N$ matrix $M^{(2)}$ has its entries given
by the somewhat simpler formula 
\[
M_{ij}^{(2)}=\ln\left(1+\frac{\sinh\left(\pi\frac{i}{j+1}\right)^{2}}{\sinh\left(\pi\frac{\epsilon_{0}}{j+1}\right)^{2}}\right).
\]
And this matrix turns out to be well conditioned numerically, even
for large $N$. 

\subsection{Interpreting the rotational data as a filtered Radon transform}

To fix ideas, let us assume that the axis of rotation of the imaging
device is the line of direction $(0,0,1)$, which passes through the
origin. Suppose the solutions of (\ref{eq:DefUoJ}), named $\left(u_{i}^{0}\right)_{1\leq i\leq N}$
correspond to $\theta=0$. After a rotation of an angle $\theta$,
they become $u_{i}^{0,\theta}(x,y)=u_{i}^{0}\left(R_{-\theta}\underline{x}\right)=u_{i}^{0}\left(\left[\begin{array}{cc}
\cos\theta & \sin\theta\\
-\sin\theta & \cos\theta
\end{array}\right]\left[\begin{array}{c}
x\\
y
\end{array}\right],z\right)$. Thus the measured data becomes
\begin{align}
\Delta C_{ij}^{\theta} & \approx\int_{\mathbb{R}^{3}}\varepsilon_{0}\left(\varepsilon_{r}\left(\underline{x}\right)-1\right)\left(\nabla u_{j}^{0}\cdot\nabla u_{i}^{0}\right)\left(R_{-\theta}\underline{x}\right)\text{d}x\text{d}y\text{d}z,\nonumber \\
 & =\int_{\mathbb{R}^{3}}\varepsilon_{0}\left(\varepsilon_{r}\left(R_{\theta}\underline{x}\right)-1\right)\left(\nabla u_{j}^{0}\cdot\nabla u_{i}^{0}\right)\left(\underline{x}\right)\text{d}x\text{d}y\text{d}z.\label{eq:Cdeltateha}
\end{align}
In particular, formula (\ref{eq:Hnonrot}) becomes 
\begin{equation}
H_{i,k}^{\theta}\approx\int_{\mathbb{R}^{3}}\varepsilon_{0}\left(\varepsilon_{r}\left(R_{\theta}\underline{x}\right)-1\right)\psi_{k}\left(\underline{x}-x_{i},z\right)\text{d}x\text{d}y\text{d}z,\label{eq:Hfinal2}
\end{equation}
with $x_{i}=\left(i-n-1\right)d\in\left\llbracket -nd,\left(n-k\right)d\right\rrbracket $
. But for the fact that the filter (or weight) $\psi_{k}$ is not constant
on its support, $H_{i,k}^{\theta}$ represents the planar Radon Transform
of the vertically averaged value of $\varepsilon_{0}\left(\varepsilon_{r}-1\right)$
on the support of $\psi_{k}$. 

\subsection{Incorporating practical sensor design constraints}

\begin{figure}
\centering{}\resizebox{0.8\columnwidth}{!}{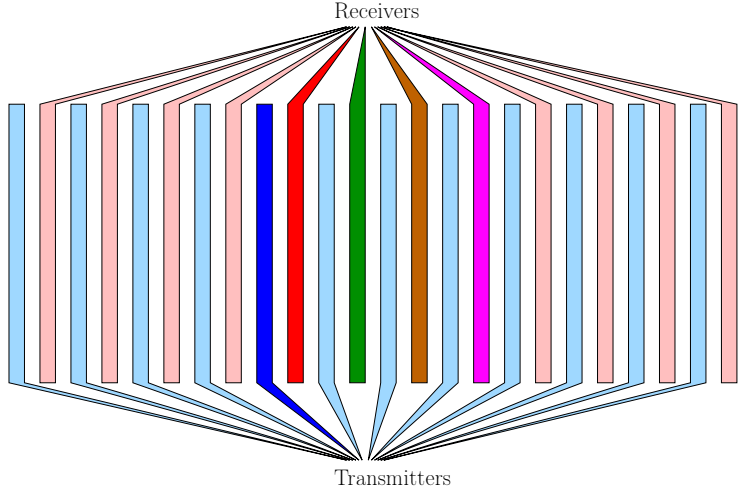}\caption{\label{fig:Sensors} Transmitting and receiving electrodes. The transmitting
electrodes are represented in blue and connected to the hardware below.
The receiving electrodes are connected to the hardware above.}
\end{figure}

The model we introduced is a simplification of the design used in
practice. The transmitting and receiving electrodes are distinct sets,
as represented in \figref{Sensors}. The receiving electrode
is grounded, whereas the non-receiving electrodes are floating, that
is, the potential $u$ satisfies a Neumann (no flux) boundary condition
on their surface. 

\begin{figure}
\begin{centering}
\includegraphics[width=0.33\columnwidth]{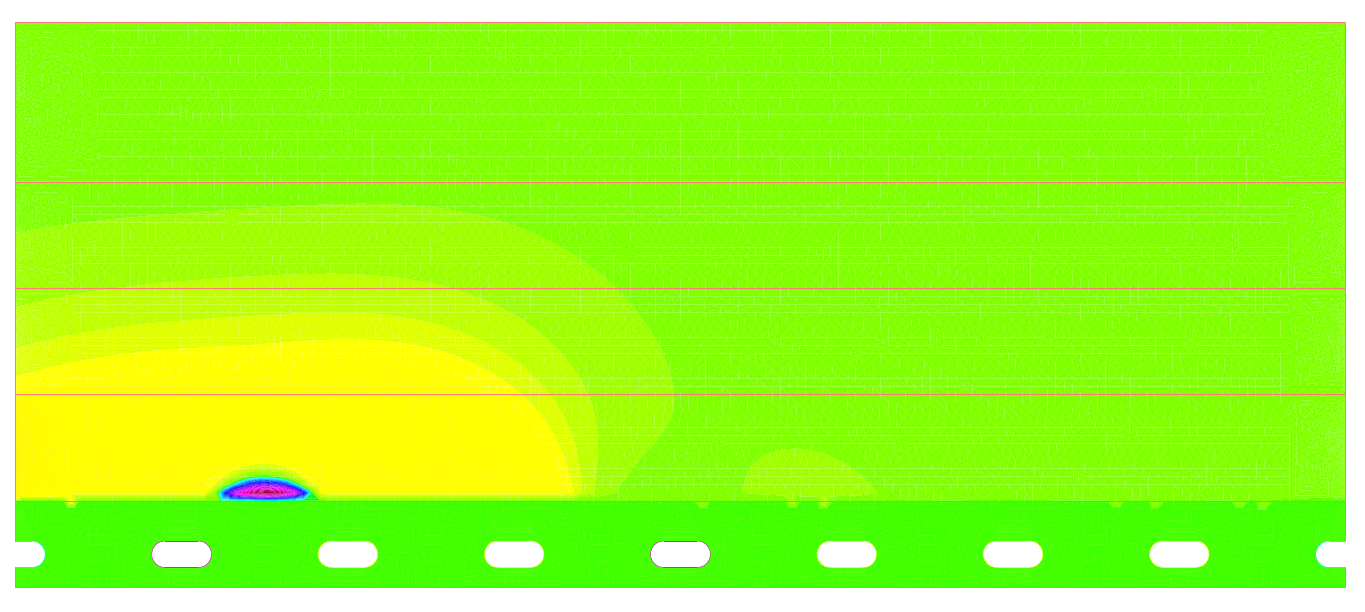}\includegraphics[width=0.33\columnwidth]{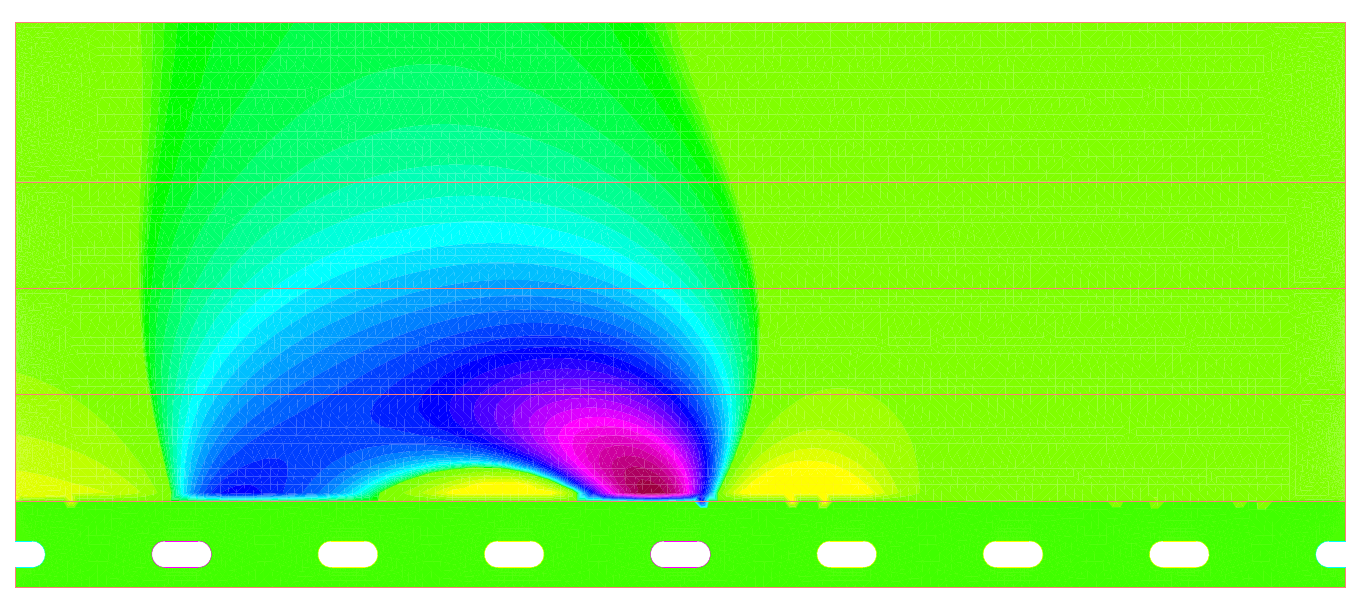}\includegraphics[width=0.33\columnwidth]{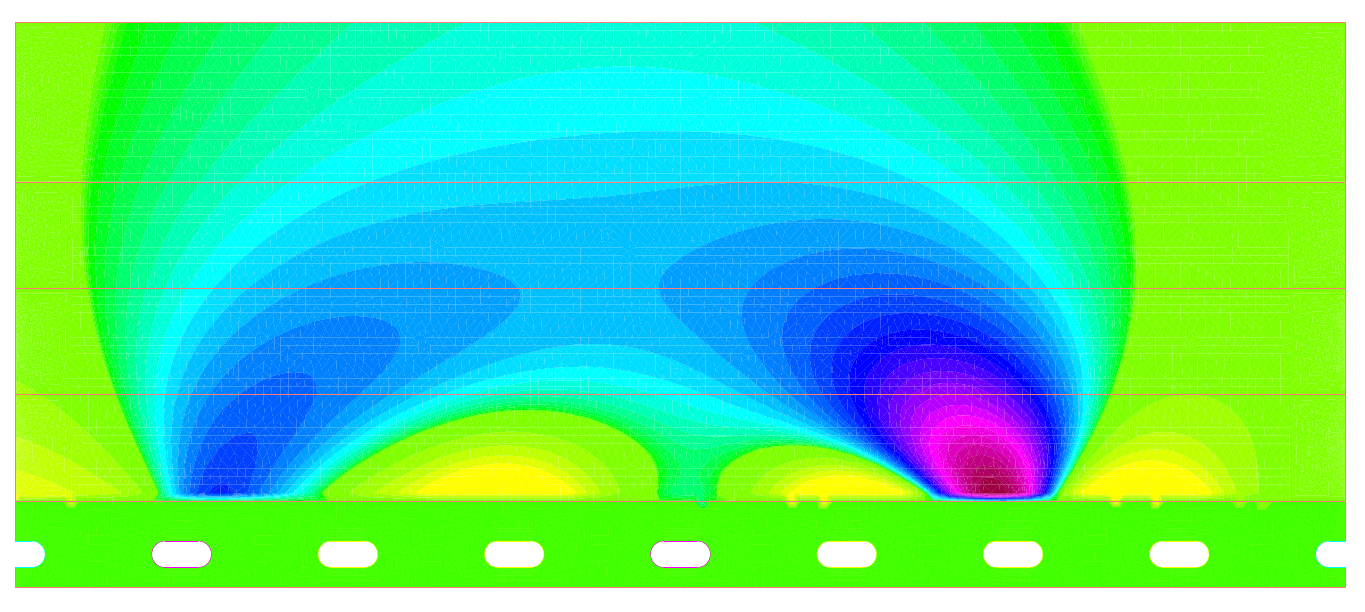}
\par\end{centering}
\centering{}\caption{\label{fig:Neumann}The first three horizontal approximate horizontal
filters $\psi_{1},\psi_{2}$ and $\psi_{3}$, taking into account
the Transmitter/Receiver distinction. Computed with FreeFem++ \cite{FreeFem}. }
\end{figure}

Such a modification changes the shape of the flux, and therefore the
definition of weight functions $\psi_{k},k=1,\ldots,4$, $\phi_{k},k=1,\ldots,4$
and $\zeta$; it does not otherwise affect the analysis in a significant
manner otherwise. To be precise (\ref{eq:Hnonrot}) still holds, but
the two variable functions $\psi_{k}$ are now given by 
\[
\psi_{k}=\nabla u_{t}^{k}(x,z)\cdot\nabla u_{r}^{k}(x,z),
\]
where for every $k$, both $u_{t}^{k}$ and $u_{r}^{k}$ satisfy different
boundary conditions -- in contrast to the previous situation where
only one potential $u_{0}$ was needed. The potentials $u_{t}^{k}$ 
and $u_{r}^{k}$ satisfy the same equation and boundary condition away 
from the electrodes, which is
\[
\begin{cases}
\partial_{x}\left(\varepsilon_{0}\partial_{x}u_{t/r}^{k}\right)+\partial_{x}\left(\varepsilon_{0}\partial_{z}u_{t/r}^{k}\right)=0 & 
\text{in }\text{\ensuremath{\mathbb{R}\times\left(0,\infty\right)\setminus\cup_{i=-\infty}^{\infty}I_{i}}}\\
\varepsilon_{0}\partial_{z}u_{t/r}^{k}\left(x_{1},0\right)=0, & \text{at }x_{2}=0\\
{\displaystyle \lim_{\left|\underline{x}\right|\to\infty}}u_{t/r}^{k}\left(\underline{x}\right)=0.
\end{cases},
\]
On the electrodes $u_{t}^{k}$ and $u_{r}^{k}$ satisfies the boundary
conditions
\[
\begin{cases}
u_{t/r}^{k}=1 & \text{on }\partial I_{0}\\
u_{t/r}^{k}=\alpha & \text{on }\partial I_{2m},\text{for every }m\in\mathbb{Z}\\
\varepsilon_{0}\partial_{\nu}u_{t/r}^{k}=0 & \text{ on }\partial I_{2l+1},\text{for all }l\in\mathbb{Z}\text{ such that }l\neq k-1\\
u_{t/r}^{k}=\beta & \text{on }\partial I_{2k-1}.
\end{cases},
\]
with $(\alpha,\beta)=(1,0)$ for $u_{t}^{k}$ 
and $(\alpha,\beta)=(0,1)$ for $u_{r}^{k}$ 
respectively.
In \figref{Neumann} we represent the functions $\psi_{1}$,
$\psi_{2}$ and $\psi_{3}$, using the same colour scale as in \figref{Psik}
and \figref{APsik}. The horizontal symmetry is broken, as we should
expect, since the boundary conditions of the Electrodes nearest the
transmitter and the receiver are different. The weight $\psi_{1}$is
the most significantly different from the others, as it takes both
positive and negative values within its support. The overall support
still roughly describes the arch intuited in the introduction, whereby
higher strata are reached by more distant electrodes. After inversion, 
the data can thus be interpreted as several 2 dimensional 
horizontal cross sections of the permittivity, at different depths -- a $2d+1$ imaging device.

\section{Experimental data}

A Radon transform inversion was performed on the experimental data obtained from the rotating imaging device. 
Here, $n=27$ (55 electrodes), and $p=180$, which means measurements were recorded after a each $1^{0}$ rotation. The
Matlab function {\tt{iradon}} was applied to the resulting data, with linear
interpolation and a Hamming filter.

\begin{figure}
\centering{}\includegraphics[width=0.95\columnwidth]{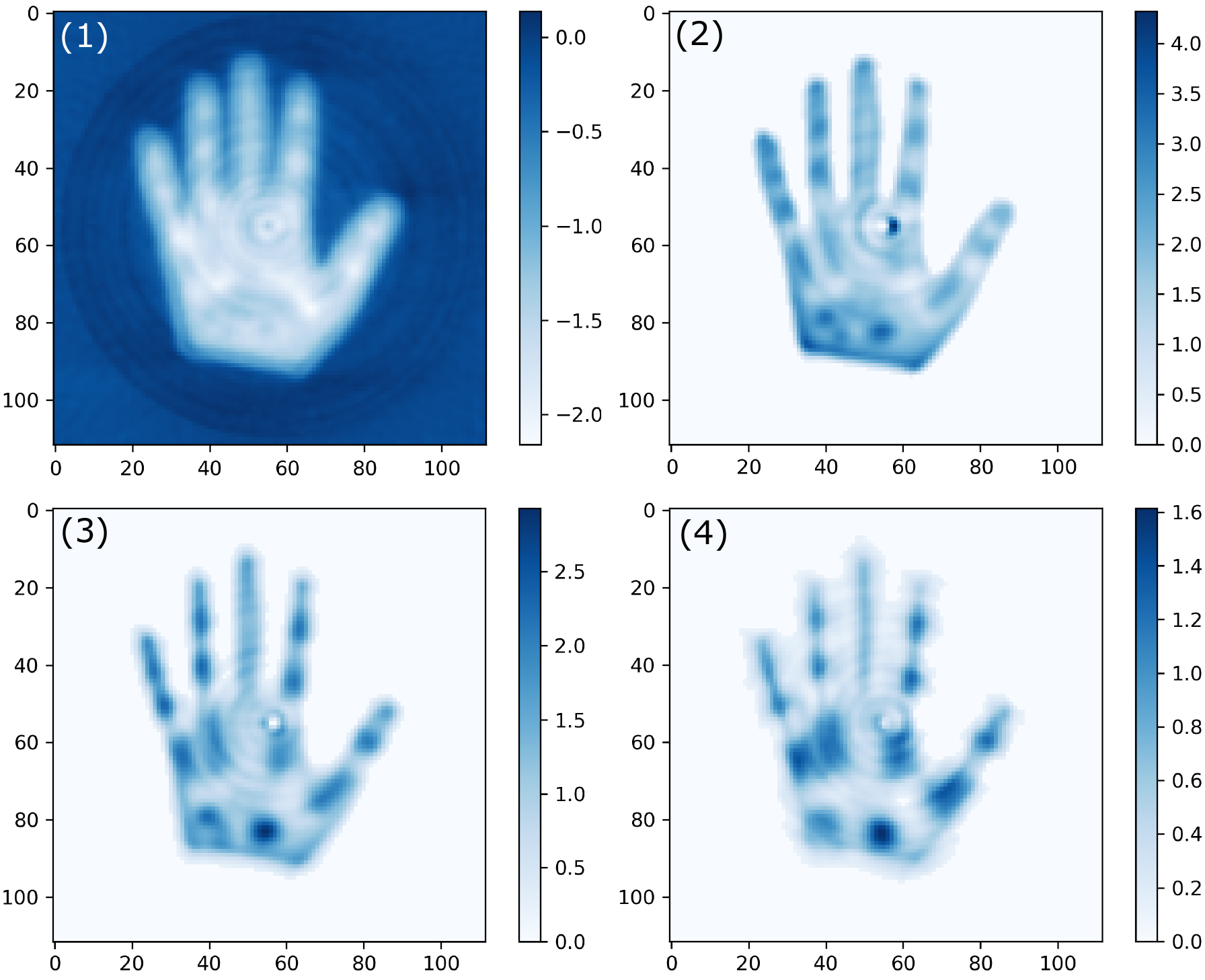}\caption{\label{fig:experimental-radon} (1 - 4) 
Images reconstructed from $H_{1}$, $H_{2}$, $H_{3}$ and $H_{4}$ (Layers 1 - 4), corresponding to data from nearest, 2nd, 3rd and 4th nearest electrode pairs.}
\end{figure}

In \figref{experimental-radon}(1) we see the result of the lowest layer, the one closest to the sensor, 
corresponding to the data $H_{1}^{\theta}(1\ldots2n)$. The shape of the hand is clearly visible. Levels of 
darkness indicate hints of lower and higher densities above. This can be expected: the field really reaches 
much more than simply the lowest level. 

In \figref{experimental-radon}(2) we show the second layer, namely, the inversion of $H_{2}^{\theta}(1\ldots2n-1)$. 
A more defined hand appears, and the silver-inked trenches appear more marked than the empty trenches. 
We report the third neighbour data inversion in \figref{experimental-radon}(3), and the fourth neighbour 
data in \figref{experimental-radon}(4). As more features are lost with depth, the trenches become more defined. 
Additionally, the trenches with missing silver ink do not appear on the images due to the high contrast caused by 
the silver ink.

\section{Conclusion, discussion, and future direction}

We have described and modelled a coplanar non invasive non destructive
capacitive imaging device. It is multiplexing, namely, individual
electrodes can be measured independently. Together with a rotating
platform, this allows to obtain Radon-like data, which, when inverted
provides very good images of the model hand at different depth. It
could seem remarkable that a Radon transform could be successfully
used for ECT, as this is not a wave propagation problem. The link
between EIT and generalized Radon Transform was established in \cite{MR1036240}
in the analysis of the Back Projection Algorithm \cite{ISI:A1984TF31700004}
and further studied in \cite{MR1104811,MR2262749}. Our pictures are
a naive double Radon inversion, namely, an actual radon inversion
in the plane direction for electrodes of increasing distance, and
then a `further away means deeper' heuristic when reading the pictures.
This calls for a proper justification, and a more quantitative algorithm,
but the pictures obtained are encouraging. The images provided are
truly raw data: no AI or compensation has been performed for the various
weights involved. The cross data measurements have not been used in
this application; they provide a convenient tool to localise horizontally
the object at hand. Here, the location of the object was clearly visible.
This data corresponds to measurements at one time point. We are investigating
what is the best way to make use of various data time points, to enrich
or improve the data collected. The quality of the images already obtained
encourages us look for optimal reconstruction methods. The advantage
of the one we have used here is that it used off-the-shelf algorithms,
and data acquisition is very fast, as only a few milliseconds are
required for each angle. \bibliographystyle{siam}
\bibliography{Zedsen}

\end{document}